

\input epsf

\input phyzzx

\overfullrule=0pt\hsize=6.5truein
\rightline{\vbox{\halign{#\hfil\cr ADP-94-17/M24\cr hep-th/9412076\cr}}}
\bigskip\bigskip\baselineskip=24pt \title{\seventeenbf Charged Dilaton Black
Holes\break with a Cosmological Constant} \baselineskip=18pt \vfill

\author{S.J. Poletti$^1$\foot{E-mail: spoletti@physics.adelaide.edu.au}, J.
Twamley$^{1,2}$\foot{E-mail: twamley@physics.uq.oz.au} and D.L.
Wiltshire$^1$\foot{E-mail: dlw@physics.adelaide.edu.au} }

\address{$^1$ Department of Physics and Mathematical Physics, University of
Adelaide,\break Adelaide, S.A. 5005, Australia.} \address{$^2$ Physics
Department, University of Queensland, St Lucia, QLD 4072, Australia.}\vfill

\abstract

The properties of static spherically symmetric black holes, which are either
electrically or magnetically charged, and which are coupled to the dilaton in
the presence of a cosmological constant, $\Lambda$, are considered. It is shown
that such solutions do not exist if $\Lambda>0$ (in arbitrary spacetime
dimension $\ge4$). However, asymptotically anti-de Sitter black hole solutions
with a single horizon do exist if $\Lambda<0$. These solutions are studied
numerically in four dimensions and the thermodynamic properties of the
solutions are derived. The extreme solutions are found to have zero entropy and
infinite temperature for all non-zero values of the dilaton coupling constant.
\vfill\centerline{(December, 1994)}\endpage
\baselineskip=16pt plus.2pt minus.2pt
\REF\GM{G.W. Gibbons and K. Maeda, \NP{B298}, 741 (1988).}
\REF\GHS{D. Garfinkle, G.T. Horowitz and A. Strominger, \PR{D43}, 3140 (1991);
(E) {\bf 45} 3888 (1992).}
\REF\DGKT{F. Dowker, J.P. Gauntlett, D.A. Kastor and J. Traschen, \PR{D49},
2909 (1994).}
\REF\mp{K. Shiraishi, \JMP{34} (1993) 1480;\br
G.T. Horowitz, in B.L. Hu and T.A. Jacobson (eds.), ``{\it Directions
in General Relativity: Proc. 1993 Int. Symp., Maryland, Vol. II$\,$}''
(Cambridge University Press, 1993), 157.}
\REF\HoHo{J.H. Horne and G.T. Horowitz, \PR{D48}, R5457 (1993).}
\REF\MS{T. Maki and K. Shiraishi, \CQG{10}, 2171 (1993).}
\REF\Ok{T. Okai, ``4-dimensional dilaton black holes with cosmological
constant'', Preprint UT-679, hep-th/9406126}
\REF\PW{S.J. Poletti and D.L. Wiltshire, \PR{D50}, 7260 (1994).}
\REF\MW{S. Mignemi and D.L. Wiltshire, \CQG{6}, 987 (1989); \PR{D46}, 1475
(1992); 
D.L. Wiltshire, \PR{D44}, 1100 (1991).}
\REF\GH{R. Gregory and J.A. Harvey, \PR{D47}, 2411 (1993).}
\REF\HH{J.H. Horne and G.T. Horowitz, \NP{B399}, 169 (1993).}
\REF\GW{G.W. Gibbons and D.L. Wiltshire, \AP{N.Y.}{167}, 201 (1986); (E) {\bf
176}, 393 (1987).}
\REF\HW{C.F.E. Holzhey and F. Wilczek, \NP{B380}, 447 (1992).}
\font\sixrm=cmr6 \font\sixi=cmmi6 \font\sixsy=cmsy6 \font\sixbf=cmbx6
\font\eightrm=cmr8 \font\eighti=cmmi8 \font\eightsy=cmsy8 \font\eightbf=cmbx8
\font\eighttt=cmtt8 \font\eightit=cmti8 \font\eightsl=cmsl8

\def\tenpoint{\def\rm{\fam0\tenrm} \textfont0=\tenrm \scriptfont0=\sevenrm
\scriptscriptfont0=\fiverm \textfont1=\teni \scriptfont1=\seveni
\scriptscriptfont1=\fivei \textfont2=\tensy \scriptfont2=\sevensy
\scriptscriptfont2=\fivesy \textfont3=\tenex \scriptfont3=\tenex
\scriptscriptfont3=\tenex \textfont\itfam=\tenit \def\it{\fam\itfam\tenit}
\textfont\slfam=\tensl \def\sl{\fam\slfam\tensl} \textfont\ttfam=\tentt
\def\tt{\fam\ttfam\tentt} \textfont\bffam=\tenbf \scriptfont\bffam=\sevenbf
\scriptscriptfont\bffam=\fivebf \def\bf{\fam\bffam\tenbf}
\setbox\strutbox=\hbox{\vrule height8.5pt depth3.5pt width0pt} \let\sc=\eightrm
\let\big=\tenbig \rm} \def\eightpoint{\def\rm{\fam0\eightrm}
\textfont0=\eightrm \scriptfont0=\sixrm \scriptscriptfont0=\fiverm
\textfont1=\eighti \scriptfont1=\sixi \scriptscriptfont1=\fivei
\textfont2=\eightsy \scriptfont2=\sixsy \scriptscriptfont2=\fivesy
\textfont3=\tenex \scriptfont3=\tenex \scriptscriptfont3=\tenex
\textfont\itfam=\eightit \def\it{\fam\itfam\eightit} \textfont\slfam=\eightsl
\def\sl{\fam\slfam\eightsl} \textfont\ttfam=\eighttt
\def\tt{\fam\ttfam\eighttt} \textfont\bffam=\eightbf \scriptfont\bffam=\sixbf
\scriptscriptfont\bffam=\fivebf \def\bf{\fam\bffam\eightbf}
\setbox\strutbox=\hbox{\vrule height7pt depth2pt width0pt} \let\sc=\sixrm
\let\big=\eightbig \rm} \def\br{\hfil\break} \def\rarr{\rightarrow}
\def\scrscr{\scriptscriptstyle}  
\def\begincaption{\medskip\openup-1\jot\eightpoint} \def\scr{\scriptstyle}
\def\endcaption{\tenpoint\openup1\jot\leftskip=0pt\rightskip=0pt}
\def\caption#1#2{\message{#1}\begincaption\leftskip=15true mm\rightskip=15true
mm\vbox{\halign{\vtop{\parindent=0pt\parskip=0pt\strut##\strut}\cr{\bf#1}\quad
#2\cr}}\endcaption} 
\def\Z#1{_{\lower2pt\hbox{$\scr#1$}}} \def\X#1{_{\lower2pt\hbox{$\scrscr#1$}}}
\def\ns#1{_{\hbox{\sevenrm #1}}}\def\Ns#1{\Z{\hbox{\sevenrm #1}}}
\def\w#1{\;\hbox{#1}\;}\def\const{\hbox{const.}}\def\ep{\epsilon}
\def\ph{\phi}\def\pt{\partial}\def\OM{\Omega}\def\dd{{\rm d}} \def\LA{\Lambda}

\def\AP#1#2{Ann.\ Phys.\ (#1) {\bf#2}} \def\PR#1{Phys.\ Rev.\ {\bf#1}}
\def\CQG#1{Class.\ Quantum Grav.\ {\bf#1}}\def\NP#1{Nucl.\ Phys.\ {\bf#1}}
\def\JMP#1{J.\ Math.\ Phys.\ {\bf#1}} \def\RN{Reissner-Nordstr\"om}

\def\g#1{{\rm g}\Z#1}  \def\V{{\cal V}} \def\Db#1{(D-#1)}
\def\pz#1{\ph\Z{#1}} \def\Qg{Q^2e^{2a\ph\X0}} \def\sgn{\w{sgn}}
\def\egph{e^{2a\ph}} \def\rH{r\X{\cal H}} \def\RH{R\X{\cal H}}
\def\th{\theta} \def\pH{\ph\X{\cal H}}
\sectionstyle{{}}\def\section#1{\par \ifnum\lastpenalty=30000\else\penalty-200
\vskip\sectionskip\spacecheck\sectionminspace\fi\global\advance\sectionnumber
by 1 {\protect\xdef\sectionlabel{\the\sectionstyle{\the\sectionnumber}} \wlog
{\string\section\space\sectionlabel}}\noindent{\caps\enspace\sectionlabel.~~#1}
\par\nobreak\vskip\headskip \penalty 30000 }
\ifx\epsfbox\UnDeFiNeD\message{(NO epsf.tex, FIGURES WILL BE IGNORED)}
\def\ifig#1#2#3{\FIG#1{#2}} 
\else\message{(FIGURES WILL BE INCLUDED)}
\def\ifig#1#2#3{\FIGNUM#1\goodbreak\midinsert\centerline{#3}%
\smallskip\centerline{\vbox{\baselineskip12pt \advance\hsize by
-1truein\eightpoint\noindent{\bf Fig.~\the\figurecount:} #2}} 
\endinsert} \def\figout{{}} \fi
\section{Introduction}

Over the past few years much interest has been focussed on the properties of
charged black holes coupled to the dilaton field, generally in a manner
dictated by the low energy limit of string theory, with a massless dilaton. A
number of black hole and related solutions have been derived. These include the
static spherically symmetric solutions [\GM,\GHS], dilatonic versions of the
$C$-metric solutions [\DGKT] which represent oppositely charged black holes
undergoing uniform acceleration, and the generalisation of the
Majumdar-Papapetrou metric which represents a collection of maximally charged
black holes in an asymptotically flat background [\mp,\HoHo]. Time-dependent
Kastor-Traschen type cosmological multi-black hole solutions have been
discussed by Horne and Horowitz [\HoHo], and by Maki and Shiraishi [\MS].
However, exact solutions have been constructed only for certain special values
of the dilaton coupling and for special powers of a Liouville-type dilaton
potential [\MS], which excluded the case that the potential is simply a
cosmological constant.

It is therefore natural to ask whether static spherically symmetric solutions
representing charged black holes coupled to the dilaton also exist in the
presence of a cosmological constant. To date this question has not been
answered although some attempts have been made to understand the problem. Okai
has examined the problem using series solutions [\Ok], and has placed limits on
the number of possible horizons. Furthermore, in a recent paper [\PW] two of us
have derived the global properties of solutions in the related model described
by the action
$$\eqalign{S=\int\dd^Dx\sqrt{-g}\Biggl\{{{\cal R}\over4}-&{1\over D-2}\,g^{\mu
\nu}\pt_\mu\ph\,\pt_\nu\ph-\V(\ph)-{1\over4}\exp\left(-4\g0\ph\over D-2\right)F
_{\mu\nu}F^{\mu\nu}\cr&-{1\over2(D-2)!}\exp\left(-4\g0\ph\over D-2\right)F_{\mu
\X1\mu\X2\dots\mu\X{\!D-2}}F^{\mu\X1\mu\X2\dots\mu\X{\!D-2}}\Biggr\},\cr}\eqn
\action$$
where the dilaton potential, $\V(\ph)$, was chosen to be of Liouville
form,\break $\V=(\LA/2)\exp\left[-4\g1\ph/(D-2)\right]$. Here $F_{\mu\nu}$ is
the field strength of the electromagnetic field, and $F_{\mu\X1\mu\X2\dots\mu\X
{\!D-2}}$ the $\Db2$-form field strength of an abelian gauge field. In [\PW] it
was shown that charged black hole solutions with a realistic asymptotic
behaviour do not exist for the Liouville-type potential\foot{Similar results
apply to the uncharged case [\MW].}.
The one exception to this result was the case $\g1=0$, where it was found that
both asymptotically de Sitter and asymptotically anti-de Sitter solutions do
exist, and that the corresponding critical point is an attractor in the phase
space. To show that black hole solutions exist one must further demonstrate
that integral curves connect these critical points to regular horizons. That is
the object of the present paper. We will demonstrate that black hole spacetimes
do exist in the case of a negative cosmological constant, $\LA$, but do not
exist if $\LA>0$. This result stands in sharp contrast to the standard \RN--de
Sitter solutions.

\section{Non-existence of black holes with a positive cosmological constant}

In order to demonstrate that black holes with a positive cosmological constant
do not exist in dilaton gravity, it is convenient to adopt the coordinates used
by Garfinkle, Horowitz and Strominger [\GHS] in their discussion of the black
hole solutions with a massless dilaton [\GM], namely
$$\dd s^2=-f\dd t^2+f^{-1}\dd r^2+R^2\dd\OM^2\Z{D-2},\eqn\coorda$$
where $f=f(r)$ and $R=R(r)$, and $\dd\OM^2\Z{D-2}$ is the standard round metric
on a $\Db2$-sphere, with angular coordinates $\th_i$, $i=1\dots D-2$.

In the present paper, as in [\PW], we will consider cases in which only $F_{\mu
\nu}$ is present with
$${\bf F}=\exp\left(4\g0\ph\over D-2\right){Q_e\over R^{D-2}}\;\dd t\wedge\dd r
\,,\eqn\elec$$
corresponding to an isolated electric charge, or else only $F_{\mu\X1\mu\X2
\dots\mu\X{\!D-2}}$ is present with components
$$F_{\hat\th\X1\hat\th\X2\dots\hat\th\X{\!D-2}}={Q_m\over R^{D-2}}\ep_{\hat\th
\X1\hat\th\X2\dots\hat\th\X{\!D-2}}\eqn\mag$$
in an orthonormal frame, which is a magnetic monopole ansatz if $D=4$. Since
the field equations are invariant under the duality transformation [\GM] $Q_e
\rarr Q_m$, $\ph\rarr-\ph$, it is convenient to define a constant
$$a\equiv\cases{+2\g0/\Db2&electric ansatz \elec,\cr-2\g0/\Db2&magnetic ansatz
\mag,\cr}\eqn\defa$$
The field equations derived from \action\ may then be written
$$\eqalignno{{2\over\Db2R^{D-2}}\left[R^{D-2}f\ph'\right]'=\;&{{\dd\V}\over{\dd
\ph}}+{aQ^2\egph\over R^{2\Db2}}\,,&\eqname\feaA\cr{R''\over R}=\;&-{4\ph'^2
\over\Db2^2}\,,&\eqname\feaB\cr{1\over R^{D-2}}\left[f\left(R^{D-2}\right)'
\right]'=\;&\Db2\Db3{1\over R^2}-4\V-{2Q^2\egph\over R^{2\Db2}}\,,&
\eqname\feaC\cr}$$
which applies both to the electric and magnetic cases, with $Q=Q_e$ or $Q=Q_m$
as appropriate. Here $'\equiv d/dr$. One further field equation follows from
\feaA--\feaC\ by virtue of the Bianchi identity.

The asymptotic properties of the solutions of these field equations were
discussed in [\PW] for potentials $\V(\ph)$ of Liouville-type. In the case of a
simple cosmological constant, $\V\equiv\LA/2$, it was demonstrated that the
only possible ``realistic'' asymptotics are de Sitter or anti-Sitter type,
depending on the sign of $\LA$. Furthermore, de Sitter asymptotics are obtained
only in the region in which the Killing vector $\pt/\pt t$ is spacelike, and
consequently any black hole solutions in such a model must possess at least two
horizons. It is quite straightforward to show that in fact there are no such
solutions. We prove the result by contradiction.

Suppose that asymptotically de Sitter solutions exist with at least two
horizons, and let the two outermost horizons be labelled $r_\pm$, with $r_-<r_
+$. The requirement of regularity at the horizon means that near $r=r_+$, $f
\propto(r-r_+)$ and $\ph(r_+)$ and $R(r_+)$ are bounded with $R(r_+)\ne0$, and
similarly for $r_-$. In the case of a cosmological constant, $\V\equiv\LA/2$,
\feaA\ then implies that at both horizons
$$\ph'f'\,\Bigr|_{r_\pm}={a\Db2Q^2\egph\over2R^{2\Db2}}\,\Biggr|_{r_\pm}\eqn
\horz$$
Since $\pt/\pt t$ is spacelike in the asymptotic region, asymptotically de
Sitter solutions must have $f'(r_-)>0$ and $f'(r_+)<0$. For the moment let us
assume that $a>0$. Then \horz\ implies that $\ph'(r_-)>0$ and $\ph'(r_+)<0$.
These two values of $\ph'$ must be smoothly connected and thus $\ph'(r)$ must
go through zero at least once in the interval $(r_-,r_+)$ at a point $r\Z0$
such that $\ph''(r\Z0)<0$. However, since $f(r)>0$ on the interval $(r_-,r_+)$
it follows from \feaA\ that if $\ph'(r\Z0)=0$ then $\sgn\ph''(r\Z0)=\sgn a>0$.
We thus obtain a contradiction. If $a<0$, then each of the signs $\ph'(r_-)$, $
\ph'(r_+)$ and $\ph''(r\Z0)$ is reversed in the argument above and we once
again obtain a contradiction. Finally, in the case of anti-de Sitter
asymptotics two horizons are also ruled out, as one must then simultaneously
change the signs of $f'(r_-)$, $f'(r_+)$ and $f(r)$ on the interval $(r_-,r_+)
$.

We note in passing that our argument is readily extended to rule out static
spherically symmetric solutions with two horizons in the case that $\V(\ph)$ is
a monotonic function with $\sgn{\dd\V\over\dd\ph}=\sgn a$. For example, in the
case of a Liouville-type potential,
$\V=(\LA/2)\exp\left[-4\g1\ph/(D-2)\right]$, such solutions do not exist if
$a/(\g1\LA)<0$. This accords with the results of [\PW], since it was found
there that Robinson-Bertotti type solutions can only exist if $a/(\g1\LA)>0$,
and it is well-known that these latter solutions only exist in the same
circumstances as solutions with two degenerate horizons. Of course, as was
observed in [\PW], the general models with a Liouville potential do not possess
realistic asymptotics.

\section{Asymptotically anti-de Sitter black holes}

Let us now turn to the case of a negative cosmological constant $\V\equiv\LA/2
$, with $\LA<0$. Since $\pt/\pt t$ is timelike in the asymptotic region, black
hole solutions with a single horizon can exist in this case, and as was shown
in the preceding section this is in fact the maximum number of horizons
possible. As a starting point, it is straightforward to determine the large $r$
behaviour of the asymptotically anti-de Sitter solutions. If we make the
expansions $\ph=\sum_{i=0}^\infty\pz ir^{-i}$, $f=\sum_{i=-2}^\infty f\Z ir^{-i
}$, $R=r+\sum_{i=0}^\infty R\Z ir^{-i}$, and furthermore use the freedom of
translating the origin in the radial direction to set $R\Z0=0$, we find
$$\eqalign{\ph=\;&\pz0+{\pz{D-1}\over r^{D-1}}+\dots\cr f=\;&{-2\LA r^2\over{(D
-1)(D-2)}}+1-{2M\over r^{D-3}}+{2\Qg\over{(D-2)(D-3)r^{2D-6}}}+{8\LA\pz{D-1}^2
\over{(2D-3)(D-2)^3r^{2D-4}}}+\dots\cr R=\;&r-{2\Db1\pz{D-1}^2\over{(2D-3)(D-2)
^2r^{2D-3}}}+\dots\cr}\eqn\assia$$
The constants $M$, $\pz0$ and $\pz{D-1}$ are free, $M$ being proportional to
the ADM mass.

One should compare these results to those of Gregory and Harvey [\GH], and
Horne and Horowitz [\HH], who investigated black holes coupled to a massive
dilaton with quadratic potential. As in the models of [\GH,\HH] the force
associated with the dilaton here is short range, but the strength of its
contribution is (for $D=4$) one power of $r$ stronger here than for the massive
dilaton. Furthermore, at this stage $\pz{D-1}$ is a free parameter, whereas it
is fixed in terms of the other charges in the massive dilaton models simply on
the basis of solving the asymptotic field equations. However, although $\pz{D-1
}$ is a free parameter as far as the asymptotic series is concerned, if we
further demand that a particular solution with an asymptotic expansion \assia\
corresponds to a black hole, then we can integrate equation \feaA\ between the
horizon, $\rH$, and infinity to obtain an integral relation
$$\pz{D-1}={a\Db2^2Q^2\over4\LA}\int_{\rH}^\infty\dd r\,{\egph\over R^{D-2}}\,.
\eqn\intrel$$
Consequently, for black hole solutions $\pz{D-1}$ is constrained to depend on
the other charges of the theory, and cannot be regarded as an independent
``hair''.

Unfortunately there is no transparent means for obtaining the general solution
to the field equations in closed form. We therefore turn to numerical
integration. Following [\HH] we will change coordinates and use $R$ as the
radial variable, so that the metric becomes
$$\dd s^2=-f\dd t^2+h^{-1}\dd R^2+R^2\dd\OM\Z{D-2}\eqn\coorb$$
where $h(R)\equiv f\left(\dd R\over\dd r\right)^2$, and now $f=f(R)$. The
advantage of working with these coordinates is that by suitably combining the
appropriate differential equations one can solve for $f$ in terms of $h$ and
$\ph$. One finds
$$f=h\exp\left[{8\over\Db2^2}\int^R\dd\bar R\,\dot\ph^2\bar R\right]\,,\eqn
\gabi$$
where $.\equiv\dd/\dd R$. There are then just two independent field equations
remaining, which (with $\V\equiv\LA/2$ and $\bf F$ corresponding to an electric
field) are
$$\eqalignno{-&R\dot h+\Db3\left(1-h\right)={4R^2h\dot\ph^2\over\Db2^2}+{2\LA R
^2\over D-2}+{2Q^2\egph\over\Db2R^{2D-6}},&\eqname\febA\cr&Rh\ddot\ph+\dot\ph
\left[D-3+h-{2\LA R^2\over D-2}-{2Q^2\egph\over\Db2R^{2D-6}}\right]={1\over2}
\Db2aQ^2{\egph\over R^{2D-5}}.&\eqname\febB\cr}$$
The asymptotic series \assia\ become
$$\eqalign{\ph=\;&\pz0+{\pz{D-1}\over R^{D-1}}+\dots\cr h=\;&{-2\LA R^2\over{(D
-1)(D-2)}}+1-{2M\over R^{D-3}}+{2\Qg\over{(D-2)(D-3)R^{2D-6}}}-{8\LA\pz{D-1}^2
\over{\Db2^3R^{2D-4}}}+\dots\cr}\eqn\assib$$
in terms of the new variables.

Since the equations are invariant under the rescaling $R\rarr c R$, $\LA
\rarr\LA/c^2$, $Q\rarr c^{D-3}Q$ it is possible to eliminate $\LA$. We will
therefore set $\LA=-1$. Furthermore, the equations are invariant under $\ph
\rarr\ph-\pz0$, $Q\rarr Qe^{a\ph\X0}$ and so one can also set $\pz0=0$ with no
loss of generality. For numerical integration it is of course necessary to
choose a particular spacetime dimension, so we will henceforth take $D=4$.

Equations \febA, \febB\ are equivalent to three first order ordinary
differential equations and thus generally have a three parameter set of
solutions. However, many of these will correspond to naked singularites. The
requirement that solutions have a regular horizon reduces the three parameters
to two, which may be taken to be the radial position of the horizon, $\RH$, and
$\pH\equiv\ph(\RH)$. Since the equations are singular on the horizon, we start
the integration a small distance from $\RH$, the initial values of $h$, $\ph$
and $\dot\ph$ being determined in terms of $\RH$ and $\pH$ by solving for the
coefficients $\tilde h_i$ and $\tilde\ph_i$ in the power series expansions, $h=
\sum_{i=1}^\infty\tilde h_i\left(R-\RH\right)^i$, $\ph=\pH+\sum_{i=1}^\infty
\tilde\ph_i\left(R-\RH\right)^i$. Since the solutions are rather cumbersome we
will not list them here.  As shown above, black hole solutions can only have
one horizon, and so we may restrict the initial conditions to those with $\dot
h(\RH)>0$.

As was shown in [\PW], the critical point of the field equations corresponding
to asymptotically anti-de Sitter solutions is a strong attractor, and thus
integrating out from a regular horizon it is possible to find solutions for
which $h(R)$ and $a\ph(R)$ increase until $h(R)$ and $\ph(R)$ eventually agree
with the asymptotic expansions \assib\ to arbitrary accuracy. The system is
thus far more amenable to numerical analysis than the corresponding system with
a quadratic dilaton potential [\HH]. Not all initial conditions with $\dot h(
\RH)>0$ lead to asymptotically anti-de Sitter solutions, however. Typically, we
find that in some instances $h(R)$ increases to a maximum, decreases to a small
positive minimum, and then finally both $h(R)$ and $|\phi(R)|$ diverge to $+
\infty$ at a finite value of $R$. Essentially, $h(R)$ comes close to displaying
a second horizon -- however, as was shown above no double horizon solutions
exist, and thus from \febB\ $\ddot\ph$ must diverge as $h$ comes close to zero
the second time. The final behaviour of such solutions, when examined in the
coordinates \coorda, is found to correspond to a central singularity.
\ifig\rawmass{Contours of constant mass of black holes as a function of the
unrescaled value of $\pH$ and $\RH$, for solutions with $Q=1$ and $a=-1$, i.e.,
magnetic solutions if $\g0=1$. For the corresponding electric solutions ($a=1$)
$\pH\rarr-\pH$ in the contour plot.}{\epsfbox{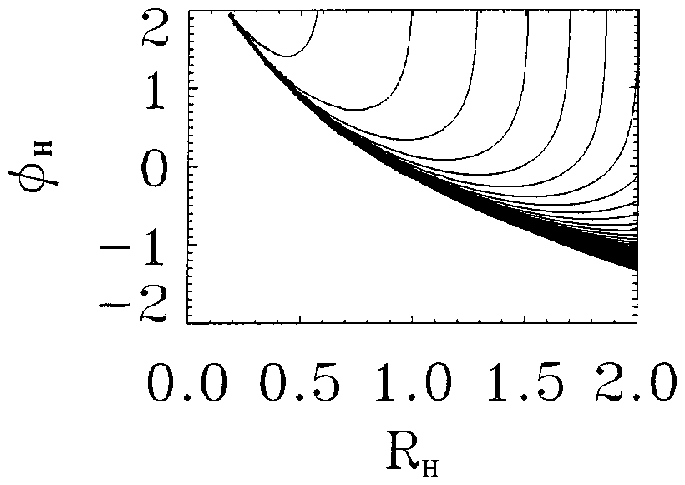}}

In order to study the properties of the solutions we performed a large number
($\gsim10^4$) of integrations. Rather than varying the charge $Q$ as an initial
condition, we adopted the approach of initially fixing $Q=1$ and integrating
out until the solutions matched the expansions \assib\ for some arbitrary
$\pz0=k$. If we then subtract the constant $k$ from the solution for $\ph$ thus
obtained, by the invariance of the equations mentioned above we have the
equivalent solutions with $\pz0=0$ and $Q=e^{ak}$. Fig.\ 1 is a contour
plot\foot{All figures are given for the case $D=4$, $\LA=-1$, and $|a|=1$.
Other non-zero values of $a$ (and other negative values of $\LA$) give results
which are qualitatively the same.} showing the integration constant $M$ as a
function of $\RH$ and $\pH$ (unrescaled). An interesting feature of the diagram
is the critical line which separates the parameter space into a region which
admits black hole solutions (on the right) from the region with no black hole
solutions. As the critical line is approached both $M$ and the effective charge
at infinity, $e^{a\ph\X0}$, become infinite. It is clear from the plot that if
$\RH$ and $-a\pH$ are both large then $M$ is almost independent of $\pH$.
However, for small $\RH$ the relation between $M$, $\pH$ and $\RH$ is quite
complicated. If the solutions are rescaled so that $\pz0=0$ as above then the
region of parameter space occupied by the black hole solutions maps into the
region $\RH>0$, $\pH{}\Ns{rescaled}>0$. To investigate the features of the
solutions we will take $M$ and $Q$ as the two independent parameters, (after
rescaling the raw numerical data), as this allows a more direct comparison with
known exact solutions than can be obtained using $\pH$ and $\RH$.
\ifig\phich{Contours of constant ``scalar charge'' $\pz3$ as a function of
mass, $M$, and charge, $Q$, for solutions with $|a|=1$. $\pz3=0$ for the
Schwarzschild-anti-de Sitter solution ($Q=0$). If $a=1$ ($a=-1$) then $\pz3<0
$ ($\pz3>0$), and $|\pz3|$ increases monotonically with increasing $Q$ for
given $M$.} {\epsfbox{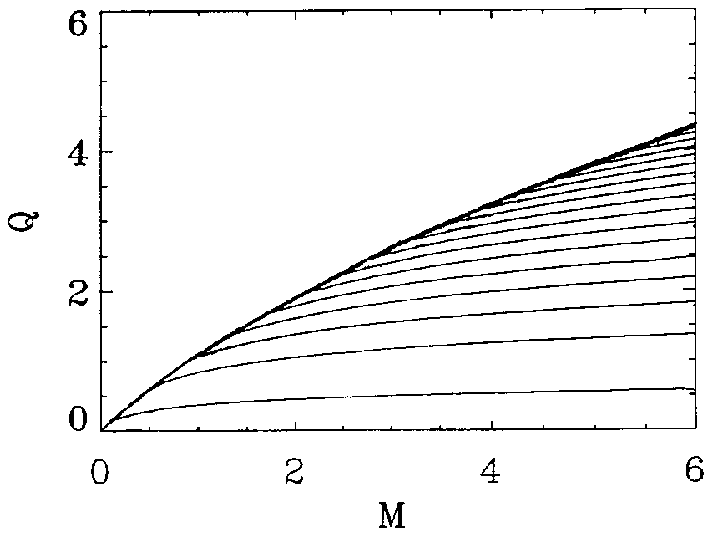}}

In Fig.\ 2 we display a contour plot of the charge $\pz3$ as a function of $M$
and $Q$, which from \assib\ (with $\pz0=0$) determines the leading order
asymptotic behaviour of the dilaton. For very large $\RH$, which effectively
means very large $M$ and small $Q$, the asymptotic series \assia\ and \assib\
are valid right up to the horizon, so that using \intrel\ we find $\pz3\approx-
aQ^2/(6M\LA)^{1/3}$, and thus $\pz3=\const$ contours follow curves $Q\propto M^
{1/6}$ as is evident in Fig.\ 2.

We have been unable to determine an analytic expression for the relation
between $Q$ and $M$ which holds for the extreme solutions, although the general
form is clear from Fig.\ 2. Since the solutions have only one horizon the
extreme black holes correspond to the singular limit $\RH\rarr0$, as in the
case of the asymptotically flat black holes with massless dilaton [\GM]. In the
case of a quadratic dilaton potential, by contrast, black holes which are large
with respect to the Compton wavelength of the dilaton have two horizons so that
the extreme limit is similar to that of the \RN\ solution and can be estimated
in the regime where the asymptotic expansions hold up to the degenerate horizon
[\HH]. In the present model, however, the extreme solutions have $\RH=0$
regardless of their mass, and so the extreme limit is closer to the situation
of the ``small black holes'' described in [\HH]. In particular, for very small
black holes with $\RH\ll(-\LA)^{-1/2}$ we expect that $Q\ns{ext}$ is somewhat
less than the limit for asymptotically flat dilaton black holes but that it
approaches this value $Q\ns{ext}\rarr\left[1+a^2\right]^{1/2}M$ as $M\rarr0$.
This is indeed borne out by the numerical analysis. The only comparison we can
make for extreme solutions with large masses is the \RN-anti-de Sitter
solution, which has two degenerate horizons with
$$3\sqrt{-2\LA}\,M=\left[\sqrt{1-4\LA Q^2\ns{ext}}-1\right]^{1/2}\left[2+\sqrt{
1-4\LA Q\ns{ext}^2}\right]$$
in the extreme limit, so that $Q\ns{ext}\approx(-\LA)^{-1/6}\left({3\over2}M
\right)^{2/3}$ for large $M$. Over the range of $M$ shown in Fig.\ 2 the
extreme limit in fact comes very close to this value for larger values of $M$,
though it does in fact eventually become less than this bound.
\ifig\adiabats{Contours of constant entropy, $S$, as a function of mass, $M$,
and charge, $Q$, for magnetic or electric solutions with $|a|=1$.}
{\epsfbox{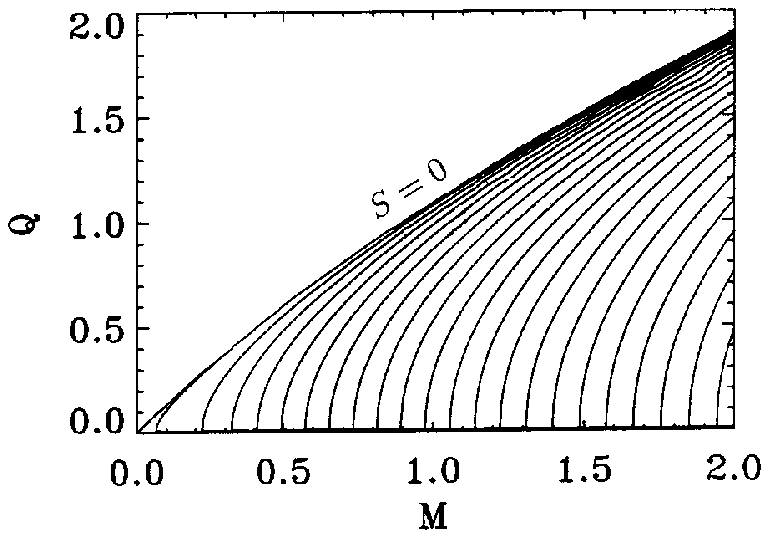}}

It is straightforward to determine the thermal properties of the black hole
solutions. The entropy is simply given by one quarter of the area of the
horizon, $S={1\over4}{\cal A}\X{\cal H}=\pi\RH^{\ 2}$. Thus for black holes
with large $M$ and small $Q$, $S\approx\pi\left(-2M\over3\LA\right)^{2/3}$,
while $S=0$ for the extreme solutions. A contour plot of the adiabats is shown
in Fig.\ 3. \ifig\isotherms{Contours of constant temperature, $T$, as a
function of mass, $M$, and charge, $Q$, for magnetic or electric solutions with
$|a|=1$. The isotherm with a temperature equal to the minimum temperature of
the Schwarzschild-anti-de Sitter solution $T\ns{cr}=\sqrt{-\LA}/(2\pi)$ is
indicated.} {\epsfbox{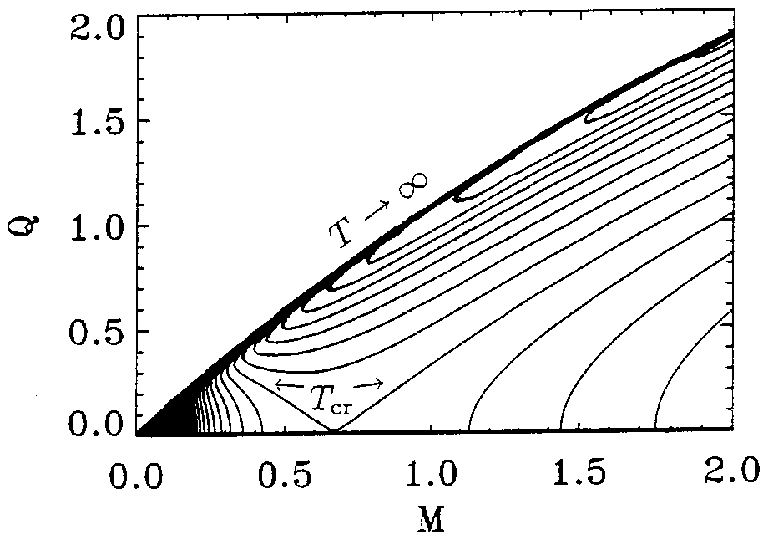}}

The temperature of the black holes is given by $T=\dot h(\RH)/(4\pi)=\tilde
h\Z1/(4\pi)$, where $\tilde h\Z1$ is determined from the power series solutions
near the horizon, giving
$$T={1\over4\pi\RH^{\ 3}}\left[\RH^{\ 2}-\LA\RH^{\ 4}-Q^2e^{2a\pH}\right]\eqn
\tempt$$
for $D=4$. Isotherm contours are plotted in Fig.\ 4. The $Q=0$ axis of course
corresponds to the Schwarzschild-anti-de Sitter solution -- the temperature for
this solution diverges at $M=0$, decreases monotonically to a minimum value $T
\ns{cr}=\sqrt{-\LA}/(2\pi)$ at $M\ns{cr}={2\over3}\left(-\LA\right)^{-1/2}$ and
then rises monotonically with increasing $M$. The isotherm $T=T\ns{cr}$
represents a critical case in Fig.\ 4 -- it has two branches, the right-hand
branch with positive specific heat, $\left(\pt M\over\pt T\right)_{\!\scrscr
Q}$, and the left-hand branch which has negative specific heat for small $Q$,
but for which the specific heat changes sign as $Q$ becomes close to the
extremal limit. Isotherms to the right of the right-hand branch have
temperatures $T>T\ns{cr}$ and the specific heat is strictly positive. Isotherms
to the left of the left-hand branch also have temperatures $T>T\ns{cr}$ and a
behaviour similar to the $T\ns{cr}$ left-hand branch, ultimately approaching
the extremal curve. In between the two $T\ns{cr}$ branches are a class of
isotherms with $T<T\ns{cr}$ which have positive specific heat for the smaller
value of $Q$ for given $M$, but which then double back with negative specific
heat close to the extremal curve. The extreme black holes have $T\rarr\infty$
for all non-zero values of $a$; this is clear from \tempt\ since the extreme
case occurs for $\RH\rarr0$ and $a\pH\rarr-\infty$.

The asymptotically flat black holes with a massless dilaton have the same
properties as the solutions here -- zero entropy, infinite temperature -- only
if $|a|>1$ [\GM,\GW]. For those solutions the temperature is zero in the
extreme limit if $|a|<1$, and finite in the intermediate `stringy' case $|a|=1$
[\GM]. Of course, an infinite temperature here merely signals the breakdown of
the semiclassical limit if one is considering the Hawking evaporation process.
As was demonstrated by Holzhey and Wilczek [\HW], in the case of the $|a|>1$
Gibbons-Maeda solutions an infinite mass gap develops for quanta with a mass
less than that of the black hole so that the Hawking radiation slows down and
comes to an end at the extremal limit, despite the infinite temperature. We
expect that the situation here would be the same.

\bigskip\noindent{\bf Acknowledgement}\quad SJP and JT would like to thank the
Australian Research Council for financial support.
\refout \figout \end